\documentstyle[graphicx,aps]{revtex}

\def\simlt{\stackrel{<}{{}_\sim}}
\def\simgt{\stackrel{>}{{}_\sim}}
\def\be{\begin{equation}}
\def\ee{\end{equation}}
\def\br{\begin{eqnarray}}
\def\er{\end{eqnarray}}
\def\NPB#1#2#3{{\it Nucl.~Phys.} {\bf{B#1}} (19#2) #3}
\def\PLB#1#2#3{{\it Phys.~Lett.} {\bf{B#1}} (19#2) #3}
\def\PRD#1#2#3{{\it Phys.~Rev.} {\bf{D#1}} (19#2) #3}

\def\PRD#1#2#3{{\it Phys.~Rev.} {\bf{D#1}} (19#2) #3}

\def\qq{q+\bar{q}}
\def\N{N_c}
\def\dgs{\delta_{GS}}
\newcommand\tev{\mathrm{TeV}}

\begin{document}
\draft
\date{August 1998}

\title{\textbf{Anomalous U(1), Gaugino Condensation and Supergravity }}
\author{T. Barreiro\footnote{e-mail : tiagob@pcss.maps.susx.ac.uk}, 
B. de Carlos\footnote{e-mail : b.de-carlos@sussex.ac.uk}}
\address{ 
Centre for Theoretical Physics \\ University of Sussex \\ Falmer, 
Brighton BN1 9QH \\ U.K.}
\author{J.~A. Casas\footnote{e-mail : casas@cc.csic.es},
J.~M. Moreno\footnote{e-mail : jmoreno@pinar1.csic.es}} 
\address{
Instituto de Estructura de la Materia \\
Consejo Superior de Investigaciones Cient\'{\i}ficas \\
Serrano 123, 28006 Madrid \\ Spain }
 
\maketitle
\begin{abstract}

The interplay between gaugino condensation and an anomalous
Fayet-Iliopoulos term in
string theories is not trivial and has important consequences
concerning the size and type of the soft SUSY breaking terms.
In this paper we examine this issue, generalizing previous
work to the supergravity context.
This allows, in particular,
to properly implement the cancellation of the cosmological
constant, which is crucial
for a correct treatment of the soft breaking terms.
We obtain that the D-term contribution to the soft masses
is expected to be larger than the F-term one. Moreover gaugino 
masses must be much smaller than scalar masses.
We illustrate these results with explicit examples.
All this has relevant phenomenological consequences,
amending previous results in the literature.

\end{abstract}

\pacs{\hfill Preprint: SUSX-TH-98-019, IEM-FT-179/98}

\newpage
 
\section{Introduction}

Superstring theories have been, for now a number of years, the most
promising candidates for physics beyond the Standard Model (SM). Two
major problems, however, have impeded extracting definite 
phenomenological predictions from these constructions: the large
vacuum degeneracy and the issue of supersymmetry (SUSY) breaking. A
large amount of work has been devoted to studying both questions,
which has led to the proposal of several mechanisms and models in
order to solve them. Among those, 
gaugino condensation \cite{deren85}
is the most
promising one, and it has been implemented, more or less successfully,
in superstring inspired scenarios. This is a non-perturbative effect
which provides the typically flat stringy fields, the
dilaton and the moduli, with a non-trivial potential which
could eventually lead to their stabilization at realistic values. 
It can also give rise to SUSY breaking at the
so-called condensation scale 
($\Lambda_c \sim M_P e^{-a/g_{string}^{2}}$ where $M_P$ is the reduced
Planck mass and $g_{string}$ is the string coupling constant), 
relating therefore both major problems of superstring phenomenology. 
So far all this is happening in the hidden sector of the theory, 
governed by a strong-type interaction, and, in this simple picture, 
gravity is responsible for transmitting the breakdown of SUSY to the 
observable sector (where the SM particles and SUSY partners live), 
parameterized in terms of the gravitino mass $m_{3/2} \sim 
\Lambda_c^3/M_P^2$ which sets the scale of the soft breaking terms.

On the other hand, it is typical of many superstring constructions to
have anomalous $U(1)_X$ symmetries whose anomaly cancellation is 
implemented by a Green-Schwarz mechanism \cite{green84,dine87,casas89}, 
where the 
dilaton plays a crucial role.
This anomalous symmetry induces a  
Fayet-Iliopoulos (F-I) D-term in the scalar potential, 
and this will generate extra 
contributions to the soft terms. 
 Therefore one might expect an interesting interplay
between SUSY breaking through gaugino condensation and the 
presence of the $U(1)_X$. 
Moreover, there is now a new scale in 
the theory, that of the $U(1)_X$ breakdown, $\xi$. Given that both 
hidden and observable sector fields are charged under this symmetry, 
the F-I term will act effectively as an extra source of transmission
of the SUSY breaking between sectors. In fact, given that in general 
$\xi < M_P$, one expects that the F-I will set the scale of the soft 
breaking terms.
 
All these issues have been recently discussed in a number of papers  
\cite{binet96,arkan98,dvali98,IbanezRoss,Ramond,JainShrock,DGPS,Chun,MohapatraRiotto,NelsonWright,FaraggiPati}, 
but always in a global SUSY limit.
In ref.~\cite{binet96} it was pointed out that the contribution
from the D-terms to the soft breaking terms is the dominant one,
which was claimed to have phenomenological merits. In particular,
this could be useful to get naturally universal soft masses, thus
avoiding dangerous flavour changing neutral current effects. 
On the other hand, in ref.~\cite{arkan98}
it was shown that that discussion had ignored the crucial role
played by the dilaton in the analysis. Furthermore, it was claimed
(based on particular ansatzs for the K\"ahler potential) that
the situation is in fact the opposite,
namely the contribution from the F-terms to the soft breaking terms
is the dominant one.

Here, we present the generalization of the results obtained in
ref.\cite{arkan98} to the supergravity (SUGRA) case, which is the
correct framework in which to deal with effective theories coming from
strings. Whereas we have checked that the SUGRA corrections 
do not affect significantly the minimization of the potential, and thus
the vacuum structure of the
theory, we have seen that they are crucial for a correct
treatment of the soft breaking terms. This is due to the fact that the
analysis of the soft terms relies heavily on the cancellation of the
cosmological constant, an issue which can only be properly addressed
in the context of SUGRA. This single extra requirement will introduce
very powerful constraints on the structure of the soft terms, which
entirely contradict previous results.

The structure of the paper is as follows: in section II we first
extend the formulation of the F-I term made in ref.\cite{arkan98} to
SUGRA, as a previous stage to analyze the vacuum structure of a typical
model first presented in \cite{binet96}. In section III we study
the cancellation of the cosmological constant, and we apply the 
constraints we get from it to the calculation of the soft breaking 
terms, obtaining a definite hierarchy between the different
contributions to the scalar masses and the gaugino mass. In section
IV we illustrate our general results with a particular example of
a gaugino condensation mechanism that stabilizes the dilaton, namely
one condensate with non-perturbative corrections to the K\"ahler 
potential. Finally, in section V we present our conclusions.

\section{Anomalous U(1), gaugino condensation and the SUGRA 
potential}

Before studying the interplay between gaugino condensation and
the presence of a Fayet-Iliopoulos D-term in string theories,
let us briefly introduce the general SUGRA formulation of such a
F-I term. As has been stressed by Arkani-Hamed et al. \cite{arkan98} 
the dilaton field, $S$ plays a crucial role in this task. Under an 
anomalous $U(1)_X$ transformation with gauge parameter $\alpha$, 
$S$ transforms as $S\rightarrow  S+i\alpha\delta_{GS}/2$, where 
the Green-Schwarz coefficient, $\delta_{GS}$, is proportional to the 
apparent $U(1)_X$ anomaly 
\be
\delta_{GS}=\frac{1}{192\pi^2}\sum_i q_i \ ,
\label{dGS}
\ee
with $q_i$  the $U(1)_X$ charges of the matter fields, $\phi_i$
(typically $|\delta_{GS}|\simlt O(10^{-1})$).
In order to be gauge invariant the K\"ahler potential must be a 
function of $S+\bar S - \delta_{GS}X$, where X is the vector
superfield of $U(1)_X$ \cite{dine87}. Therefore, the $U(1)_X$ invariant 
$D$-part of the SUGRA action reads
\be
{\cal L}_D = \int d^4 \theta \left[-3e^{-K/3}\right]
\label{LD}
\ee
with
\be
K=K(\phi^*_i e^{2q_iX}, \phi_i;\ S+\bar S - \delta_{GS}X) \;.
\label{KX}
\ee
Extracting the piece proportional to $D_X$ in Eq.~(\ref{LD}), and
eliminating $D_X$ through its equation of motion we find $D_X= g_X^2 
e^{-K/3}\left[-K_{\bar i}q_i\phi^*_i +K' \delta_{GS}/2\right]$, 
where\footnote{The convention for the sign of $D_X$ is as in 
ref.~\cite{arkan98}.} 
$g_X$ is the gauge coupling, the prime indicates a derivative with 
respect to the dilaton $S$ and $K_{\bar i}$ indicates derivatives of
the K\"ahler potential with respect to $\phi^*_i$.
Consequently, the contribution to the potential is
\be
V_D=\frac{1}{2g_X^2} D_X^2 = \frac{1}{2}g_X^2
 \left[K_{\bar i}q_i\phi^* + \xi^2 \right]^2
\label{VD}
\ee
with
\be
\xi^2= -\frac{K' \delta_{GS}}{2} \;.
\label{xi}
\ee
In writing Eq.~(\ref{VD}) we have absorbed the $e^{-2K/3}$ factor into the
redefinition of the vierbein $e\rightarrow e^{2K/3} e$, as it is 
usually done in SUGRA theories to get a standard gravity action. 
Eqs.~(\ref{VD},\ref{xi}) were obtained in ref.~\cite{arkan98} in the 
global SUSY picture. The expression of $\xi^2$ remains as given in
ref.~\cite{arkan98}.

Let us now turn to the gaugino condensation effects, and how are
they affected by the presence of the F-I potential. In order to discuss
this issue, let us consider in detail the model presented initially by
Bin\'etruy and Dudas \cite{binet96} and recently reanalyzed by 
Arkani-Hamed et al. \cite{arkan98}. The starting point is a scenario 
with gaugino condensation in the hidden sector of the theory,
originated by a $SU(N_c)$ strong-like interaction. Following 
\cite{arkan98} we take the number of flavours $N_f=1$, which 
corresponds to chiral
superfields $Q$, $\bar{Q}$ that transform under 
$SU(N_c) \times U(1)_X$ as ($N_c$,$q$) and ($\bar{N}_c$,$\bar{q}$) 
respectively. The spectrum in 
this sector is completed by a $SU(N_c)$ singlet, $\varphi$, which has 
charge $-1$ under the $U(1)_X$. It is also assumed that $\xi^2$
has positive sign.
The superpotential of this model is 
given by
\be
W = m \frac{t^2}{2} \left( \frac{\varphi}{M_P} \right)^{q+\bar{q}} + 
(N_c-1) \left( \frac{2 \Lambda^{3N_c-1}}{t^2} \right)^{\frac{1}{N_c-1}}
\;\;, \label{sup}
\ee
where $t=\sqrt{2 Q \bar{Q}}$ is the meson superfield and $\Lambda$ is
the condensation scale, which is related to the dilaton by 
\be
\label{Wnp}
\left( \frac{\Lambda}{M_P} \right)^{3N_c-1} = 
e^{-2(q+\bar{q})S/\delta_{GS}} \;\;.
\ee
As for the K\"ahler potential, $K$, it was assumed in 
\cite{binet96,arkan98} that it consists of a dilaton dependent part 
plus canonical terms for $t$ and $\varphi$. The scalar potential for 
such a theory in the framework of SUGRA is given by
\be
V = e^{K/M_P^2} \left\{ \frac{1}{K''} \left| W'+K'\frac{W}{M_P^2} 
\right|^2 + \left| W_{\varphi} + K_{\varphi} \frac{W}{M_P^2} \right|^2 
+ \left| W_t+K_t\frac{W}{M_P^2} \right|^2 - 3 \left| \frac{W}{M_P} 
\right|^2 \right\} + \frac{1}{2g_X^2} D_X^2 \;\;,
\label{pot}
\ee
where, as before, a prime indicates a derivative with respect to $S$, 
and the subindices indicate derivatives of the superpotential and 
K\"ahler potential with respect to the corresponding fields.
The terms generated by the SUGRA corrections are indicated explicitly 
by the presence of inverse powers of the Planck mass, so that in the 
limit $M_P \rightarrow \infty$ we recover the global SUSY case studied
by the authors of refs.~\cite{binet96} and \cite{arkan98}. 

The D-part of the potential reads
\be
\frac{1}{2g_X^2} D_X^2 = \frac{1}{2}g_X^2
 \left[\frac{q+\bar q}{2}|t|^2 - |\varphi^2| + \xi^2 \right]^2 \;.
\label{VD2}
\ee
This model leads to SUSY breaking by gaugino condensation
(provided that $S$ is stabilized at a non-trivial value). 
In order to study the phenomenological implications of the
presence of the F-I term, we have to compute the soft breaking terms,
separating the F-I contribution. The first step in this task is to 
minimize the potential. Following ref.~\cite{binet96} we define the 
parameters
\br
\hat{m} & = & m \left(\frac{\xi}{M_P} \right)^{q+\bar{q}} \;\;, 
\nonumber \\
& & \label{exp} \\ 
\epsilon & = & \left( \frac{\Lambda}{\xi} \right)^{\frac{3N_c-1}{N_c}}
 \left( \frac{\xi}{\hat{m}} \right)^{\frac{1}{N_c-1}} \;\;, \nonumber 
\er
with $\epsilon \ll 1$. We have checked that the expansion in $\epsilon$ 
presented by refs.~\cite{binet96,arkan98}, around $<\varphi^2>=\xi^2$ 
and $<t^2>=0$, is still correct in the SUGRA case. To be more precise, 
we will parameterize
\br
<\varphi^2> & = & \xi^2 [1+ \epsilon (q+\bar{q}) + a \; \epsilon^2 + 
...] \;\;, \nonumber \\
& & \label{sols} \\
<t^2> & = & 2 \xi^2 \epsilon [1+ b \; \epsilon + ...]
\;\;. \nonumber 
\er
It is straightforward to check that the minimization of $V$ with 
respect to the $\varphi$, $t$ fields at lowest order in $\epsilon$ 
imposes the form of the lowest order terms in Eq.~(\ref{sols}). In
order to evaluate the next to leading order coefficients in this 
expansion, $a$ and $b$, we have to solve the minimization conditions 
to the next order in $\epsilon$. For this matter, and future 
convenience, the following (lowest order in $\epsilon$) expressions 
are useful
\br
W & = & \hat{m} \xi^2 \epsilon \; N_c \;\;, \label{W} \\   
D_X & = & g^2 \xi^2 \epsilon^2 \; (a -(\qq) b) \label{DX}\;\;, \\
\label{Wt}
\frac{\partial W}{\partial t} & = & \hat{m} \xi \epsilon^{3/2}\;
\sqrt{2} \left [ \frac{(\qq)^2}{2} +  b \; \frac{N_c}{N_c-1} \right ]\;\;,
\\
\frac{\partial W}{\partial \varphi} & = & \hat{m} \xi \epsilon \; (\qq)\;\;,
\\
\frac{\partial W}{\partial S} & = & - \hat{m} \xi^2  \epsilon \;
\frac{2 (\qq)}{\delta_{GS}} \;\;.
\er
Now, from the minimization condition, $\partial_t V = 0$, we get 
\br
b  & = & b_0  + \frac{(N_c-1)}{2 K''} \left( \frac{\qq}{N_c}
\frac{K' (2 K' - \delta_{GS} K'')}{2} + \frac{K' (2 +\qq) 
\delta_{GS} K''}{2} \right) \;\;,
\er
where $b_0$ is the result obtained in \cite{arkan98} in the global SUSY case
\br
b_0  & = & \frac{(N_c-1)}{2 K''} \left( \frac{(\qq)^2}{N_c^2}
\frac{(1-2N_c) \delta_{GS}K'' - 2  K'}{\delta_{GS}} \right) \;.
\er
From the second minimization condition, $\partial_\varphi V = 0$,  
we get
\be
a  =  (\qq) \; b { \displaystyle \; - \; \frac{e^K}{g^2}
\frac{2 \hat{m}^2 (\qq)^2 } { \delta_{GS} K'} \left( 1 - 
\frac{\qq}{N_c} \left(  1-\frac{2 K'}{\delta_{GS}K''} \right)
+ \Delta_{SUGRA} \right)} \;\;,
\ee
where
\br
\Delta_{SUGRA}  = K' \left( \delta_{GS} - \frac{2 K'}{K''} \right)
- \frac{N_c}{(\qq)^2} \frac{\delta_{GS} K'^2}{4 K''} \left(
- 2 (\qq) K' + (N_c+\qq)  \delta_{GS} K''\right)   
- \frac{N_c}{\qq} \frac{3 \delta_{GS} K'}{2} \;\;,
\label{delta}
\er
to be compared to the global case
\be
a_0  =  (\qq) \; b_0 { \displaystyle - \frac{1}{g^2}
\frac{2 \hat{m}^2 (\qq)^2}{\delta_{GS} K'} \left( 1 - 
\frac{\qq}{N_c} \left( 1-\frac{2 K'}{\delta_{GS}K''} \right)
\right) } \;\;.
\ee
From the previous expressions we can write the final form of the 
D-term
\be
\label{DD}
D_X = e^K \hat{m}^2 \epsilon^2 (\qq)^2 \left( 1 - \frac{\qq}{N_c}
\left( 1-\frac{2 K'}{\delta_{GS}K''} \right) + \Delta_{SUGRA}
\right) \;\; ,
\ee
where $\Delta_{SUGRA}$ was given in Eq.~(\ref{delta}).

Finally, the third minimization condition, $\partial_S V = 0$ 
translates into an equation for the value of the third derivative of 
the K\"ahler potential, 
\br
\label{K'''}
\frac{K'''}{K'} & = &
-\frac{2 (\qq)}{\N \dgs}\frac{K''}{K'}\left( 1 
                          - \frac{\dgs K''}{2 K'} \right)^2 
+ 2 \frac{\N \dgs K' (\N \dgs K' + \qq)}{(\N\dgs K' - 2 (\qq))^2} 
                    \left( \frac{K''}{K'} \right)^2 \nonumber\\
 & & 
 -\frac{\dgs \left(\N \dgs K' (\N \dgs K' + 6 (\qq)) - 4 (\qq)^2 \right)}{
        2 (\N\dgs K' - 2 (\qq))^2}
        \left( \frac{K''}{K'} \right)^{3}.
\er
For phenomenological consistency we will assume that $K'''$ satisfies
(\ref{K'''}) at ${\rm Re} S\simeq g_{GUT}^{-2}\simeq 2$.
Notice that, since $\dgs$ is small, one expects
\be
\label{K3K2}
K'''\gg K''\ \ ,
\ee
as was already pointed out in ref.~\cite{arkan98}. In that reference it
was also claimed that one expects $K'\gg K''$, based on a
particular ansatz for $K$. This was crucial to get the result that the
F-term contribution (in particular the ${F}_S$ one) to the soft terms
dominates over the D-term one. However, as we shall shortly see, this is
not a consequence of the minimization and, in fact, general arguments
indicate that the most likely case is  precisely the opposite.

\section{Cancellation of the cosmological constant and size of the 
soft terms}

In order to get quantitative results for the soft terms we need, 
beside the above minimization conditions, an additional condition, 
which is provided by the requirement of a vanishing cosmological constant. 
Notice from 
Eqs.~(\ref{W}--\ref{Wt}) that, for this matter, the D-part of the 
potential and the $ \left| W_t + K_t W \right|^2 $ term in Eq.~(\ref{pot}),
being of order $O(\epsilon^4)$ and $O(\epsilon^3)$ respectively, are
irrelevant. Consequently, at order $O(\epsilon^2)$ the 
cancellation of the cosmological constant reads
\be
V \equiv e^K \left\{ \frac{1}{K''} \left| W'+K'{W} \right|^2 + 
\left| W_{\varphi} + K_{\varphi} {W} \right|^2 - 3 \left|W 
\right|^2 \right\} \; + \; O(\epsilon^3) \ = \ 0 \;\;.
\label{cosmo}
\ee
From this equation it can be already  noticed at first sight that
$K'$ cannot be much larger than $K''$, otherwise the first term
above cannot be cancelled by the $-3 |W|^2 $ term.
Actually, from (\ref{cosmo}) it is possible to get non-trivial bounds on the
relative (and absolute) values of $K',K''$ and thus on the relative
size of the various contributions to the soft terms. It is important
to keep in mind that $\xi^2= -{K' \delta_{GS}}/{2}$ and that
both $-K',K''$ have positive sign (the former by assumption, the latter
from positivity of the kinetic energy). Eq.~(\ref{cosmo})
translates into
\be
f(\xi^2) \left( \frac{\qq}{\N} +\xi^2\right)^2 = 3 \xi^2 \;\;,
\label{fsffi}
\ee
where 
\be
f(\xi^2) \equiv 1+ \frac{4 \xi^2}{K'' \dgs^2} \;\;.
\label{fxi}
\ee

The $1$ in the right hand side corresponds to the $|F_{\varphi}|^2
=e^K \left| W_{\varphi} + K_{\varphi} {W} \right|^2$
contribution while the other term comes from the $|F_S|^2
= e^K (K'')^{-1}\left| W'+K'{W} \right|^2$ contribution. 
Note that  $f(\xi^2) \geq 1$, reflecting the fact that both
contributions are positive definite. We can treat Eq.~(\ref{fsffi})
as a quadratic equation in $\xi^2$ which has $f(\xi^2)$-dependent 
solutions given by
\be
\xi^2 = \frac{3}{4 f(\xi^2)} \left( 1 \pm 
         \sqrt{1 - \frac{4 f(\xi^2)}{3} \frac{\qq}{N_c}} \right)^{2}
 \;\;.
\label{fsffi2}
\ee
The existence of solutions requires a positive square root.
Hence,  $f(\xi^2)$ is constrained to be within the range
\be
1 \leq f(\xi^2) \leq \frac{3 \N}{4 (\qq)} \;,
\label{bounds}
\ee
which in particular implies that
\be\label{qbound}
\frac{\qq}{\N} \leq \frac{3}{4} \;\;.
\ee
Actually, $(\qq)/\N$ is also bounded due to phenomenological
reasons. Namely $m_{3/2}\sim |W|/M_P^2$ should be $O$(1 TeV) to
guarantee reasonable soft terms. Since the perturbative and
non-perturbative contributions to $W$ (see Eq.~(\ref{sup})) are
of the same size at the minimum, we may apply this condition
to the non-perturbative piece, i.e.
\be
\label{Westim}
|W_{np}| = (N_c-1) \left(
\frac{2}{t^2}
e^{- 2(\qq) S/\dgs} \right)^{1/(\N - 1)}
\sim O(1\ \tev) \;\;,
\ee
where we have used Eq.~(\ref{Wnp}) and everything is expressed in 
Planck units. Using $t^{ -2/(N_c-1) } \simgt O(1)$ we get
${2(\qq) {\rm Re} S}/ {\delta_{GS} (N_c-1)}\simgt 34$.
Since ${\rm Re} S\simeq 2$,
we finally get
\be
\label{estimate}
\frac {\qq} {\N} 
\sim
9 \dgs\ \;\;.
\ee
Now, a non-trivial result about the relative sizes of $K'$ and $K''$ 
can be derived from
the right hand side of the Eq.~(\ref{bounds}). By writing $f(\xi^2)$
explicitly in terms of these derivatives of the K\"ahler potential
and using Eq.~(\ref{estimate}) we get
\be
\left| \frac{18 K'}{K''} \right| \leq \frac{3}{4} - 
\frac{\qq}{\N} \leq \frac{3}{4}\;,
\ee
which results in the general bound
\be
\left | \frac{K'}{K''} \right|  \leq \frac{1}{24} \;\;.
\label{kpkpp}
\ee
This bound means in particular that the $K''\ll |K'| $ assumption made
in ref.~\cite{arkan98} is clearly inconsistent with the cancellation
of the cosmological constant. As mentioned at the end of section II,
this assumption was crucial for the results obtained in that paper
concerning the relative size of the F and D contributions to the soft 
terms. This suggests that those results must be revised, as we are
about to do.

Besides Eq.~(\ref{kpkpp}), Eq.~(\ref{fsffi}) provides interesting 
separate constraints on $K',K''$. Namely, from Eq.~(\ref{fsffi}) we can
write the inequality $( \frac{\qq}{\N}+\xi^2)^2 \leq 3 \xi^2 $,
which implies
\be
\frac{3}{4} \left( 1 - 
\sqrt{1 -\frac{4}{3} \frac{\qq}{\N}} \right)^{2}
\leq \xi^2 \leq\ \frac{3}{4} \left( 1 + 
\sqrt{1 -\frac{4}{3} \frac{\qq}{\N}} \right)^{2} \;\;.
\label{ffi}
\ee
This translates into an allowed range for $K'$, since 
$\xi^2= -{K' \delta_{GS}}/{2}$.
On the other hand, Eq.~(\ref{fsffi}) also implies
the inequality $\frac{4 \xi^2}{K'' \dgs^2}
( \frac{\qq}{\N}+\xi^2)^2 \leq 3 \xi^2 $ and, thus
\be
K''\geq \frac{4}{3} \left( \frac{\qq}{\N \dgs} \right)^{2}
\sim O(100) \;\;.
\label{KSS}
\ee
Let us notice that the bounds Eq.~(\ref{ffi}) and Eq.~(\ref{KSS})
are a consequence of imposing that neither the $|F_{\varphi}|^2$
nor the $|F_{S}|^2$ contributions (both positive)
may be larger then the (negative sign) contribution $3|W|^2$
in Eq.~(\ref{cosmo}).

It is also interesting to discuss in which cases  the $|F_{S}|^2$
contribution dominates over the $|F_\varphi|^2$ one or vice-versa.
The $|F_{S}|^2$ contribution is maximized when $f(\xi^2)$ is
as large as possible, i.e. when the upper
bound in Eq.~(\ref{bounds}) gets saturated. Then
$\xi^2 = (\qq)/\N$ and hence (see Eqs.~(\ref{xi}, \ref{estimate}))
\be
\label{Kp18}
|K'| \sim 18 \;\;.
\ee
If $|K'|$ is smaller (larger) than 18, $|F_\varphi|^2$ becomes
dominant and $\xi^2$ tends to left (right) hand limit of (\ref{ffi}).
Condition (\ref{Kp18}) does not guarantee that the $|F_{S}|^2$
contribution {\em dominates} over the $|F_\varphi|^2$ one.
This would require $f(\xi^2)\gg 1$, i.e. $\left| -K'/K'' \right| 
\gg  \dgs/2$ and, from Eq.~(\ref{bounds}), $(\qq)/\N \ll 3/4$. The 
latter condition clearly shows that $|F_{S}|^2$ dominance can only 
happen in a very restricted region of parameter space, and therefore 
is unlikely to appear in explicit constructions.

Let us turn now to the important issue of the soft terms, and how
are they constrained by the previous bounds. The soft mass of
any matter field, $\phi$, is given by
\be 
\label{mfi}
m_{\phi}^2 = m_F^2 + m_D^2 \;\;,
\ee
where
\be 
\label{mFmD}
m_F^2=e^K |W|^2 \hspace{1cm} {\rm and} \hspace{1cm} m_D^2=- q_\phi <D>
\ee 
are the respective contributions from the F and D terms to $m_\phi^2$.
Here $q_\phi$ is the $\phi$ anomalous charge and we have assumed
a canonical kinetic term for $\phi$ in the K\"ahler potential
(as for $t$ and $\varphi$). From Eq.~(\ref{DD})
\be 
\label{Dsoft}
<D_X>= O(1)\times  e^K \hat{m}^2 \epsilon^2 (\qq)^2 \;\;.
\ee 
So, using Eq.~(\ref{W}) we get
\be
\label{mDmF}
\frac{m_D^2}{m_F^2}  = O(1)\times 
q_\phi \left(\frac{\qq}{N_c}\right)^2
\left(\frac{M_P}{\xi}\right)^4
\simeq O(1)\times  \left(\frac{18}{K'}\right)^2 \;.
\ee
This means that for $|K'|<18$ the D contribution to the soft masses 
will be the dominant one (contrary to what was claimed in 
ref.~\cite{arkan98}). The $|K'|>18$ case cannot be excluded, but we 
could not implement it with the explicit example we use in the next 
section. On the other hand, notice that, since 
$\xi^2\propto |K'|$ (see Eq.~(\ref{xi})), the F-I scale
becomes in this case comparable to $M_P$ (or larger).
So it is not surprising that for large $K'$ the role of gravity as the
messenger of the (F-type) SUSY breaking is not overriden
by the ``gauge mediated'' (D-type) SUSY breaking associated
to the F-I term. 

Concerning gaugino masses, these are given by
\be
\label{gaugino}
m_{\lambda}^2 = \frac{F_S^2}{(S+\bar{S})^2} \simeq 
\frac{1}{16}(K'')^{-2} e^K | W' + K'W|^2 \;\;.
\ee
Using $(K'')^{-1} | W' + K'W|^2\leq 3|W|^2$ (to allow
$V=0$ in Eq.~(\ref{cosmo})) we get
\be
\label{gaugino2}
m_{\lambda}^2 \simlt \frac{3}{16}(K'')^{-1} e^K |W|^2
=\frac{3}{16}(K'')^{-1}m_F^2 \;\;.
\ee
Since $K''\geq O(100)$, the gaugino masses are strongly suppressed
with respect to the scalar masses, which poses a problem of naturality.
Namely, since gaugino masses must be compatible with their experimental 
limits, the scalar masses must be much higher than 1 TeV, leading
to unnatural electroweak breaking. This conclusion seems inescapable 
in this context.

Summarizing, the hierarchy of masses we expect is 
\be
\label{hierarchy}
m_{\lambda}^2 \ll m_F^2 < m_D^2 \;\;,
\ee
although the last inequality might be reversed in special cases.

\section{Explicit examples}

In this section we want to illustrate the points we have just been 
discussing with a particular example. For that purpose we shall take
a model of gaugino condensation in which the dilaton is stabilized by
non perturbative corrections to the K\"ahler potential 
\cite{shenk90,banks94,casas96,barre98};
in particular the ansatz we shall use is 
\be 
K = -\log(2 {\rm Re} S) + K_{np} \;\;,
\ee
where \cite{barre98} 
\be
K_{np} = \frac{D}{B\sqrt{{\rm Re} S}} 
\log\left( 1+e^{-B (\sqrt{{\rm Re} S}-\sqrt{S_0})} \right) 
\;\;.
\label{ours}
\ee
This function depends on three parameters, $S_0$, $D$, and $B$, the 
first of which just determines the value of ${\rm Re} S$ at the minimum.
Since ${\rm Re} S\simeq 2$, we shall fix 
$S_0=2$ from now on.
Therefore this description is effectively made in terms of only $D$ 
and $B$, which are positive numbers. 

The ansatz Eq.~(\ref{ours}) can naturally implement the hierarchy
$K' \ll K''\ll K'''$, which, as we have seen in previous sections
(cf. Eqs.~(\ref{K3K2}, \ref{kpkpp}) ), is required to stabilize the 
dilaton at zero cosmological constant. Actually, there is a 
curve of values of $D$ vs. $B$ for which this happens.
This is shown in Fig.~1 for a few different 
values of $(q+\bar{q})/N_c$ ($\dgs$ being determined by
imposing a correct phenomenology); all of them below the upper
bound of $3/4$ found in the previous section, see Eq.~(\ref{qbound}).

%%%%%%%%%%%%%%%%%%%%%%%%%%%figure%%%%%%%%%%%%%%%%%%%%%%%%%%%%%
\begin{figure}
\centerline{  
\includegraphics[height=9cm]{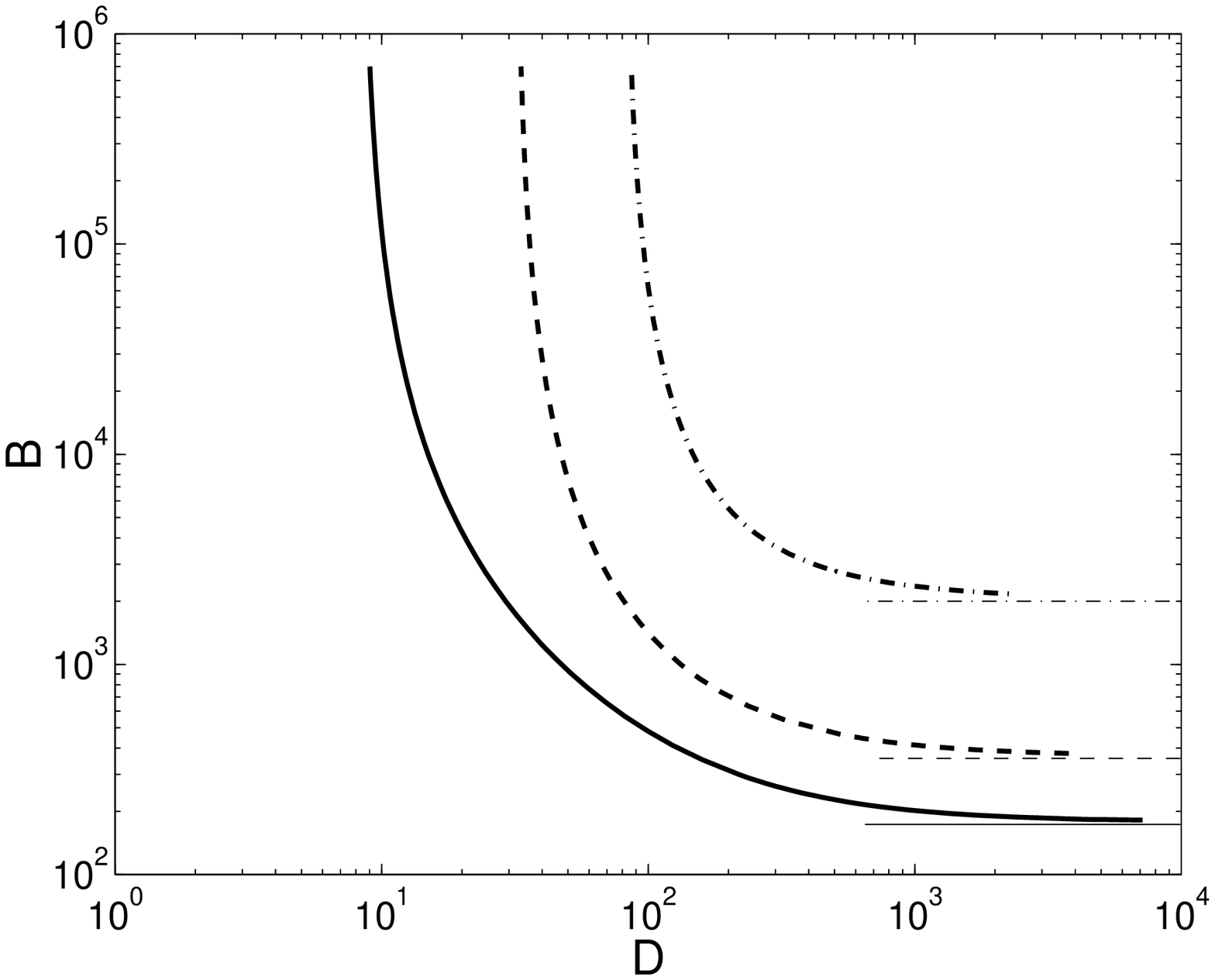}
}
\caption{}
{\footnotesize Values of the $B$ and $D$ parameters (in logarithmic 
scales) that give a zero cosmological constant with the ansatz of 
Eq.~(\ref{ours}). The hidden gauge group is SU(5), and
$m=10^{-10} M_P$ for all the curves. The solid line corresponds to 
$q + \bar{q} = 1$ and 
$\delta_{GS} = 45/192 \pi^2$, the dashed line to $q + \bar{q} = 2.5$ 
and $\delta_{GS} = 120/192 \pi^2$ and the dash-dotted line to $q + 
\bar{q} = 3.5$ and $\delta_{GS} = 150/192 \pi^2$. The thin lines show 
the corresponding lower bound on $B$ obtained from 
Eq.~(\ref{lowbbound}).}
\end{figure}
%%%%%%%%%%%%%%%%%%%%%%%%figure%%%%%%%%%%%%%%%%%%%%%%%%

The sizes of $K'$ and $K''$ at the minimum are
\br
|K'| & \sim & O(10) < 18 \nonumber \\
 & & \\
|K''| & \sim & O(100) |K'| \nonumber \;\;.
\er
The second equation satisfies the bound Eq.~(\ref{kpkpp}), while the
first one has implications on the size of the soft terms as
we shall shortly see.

Concerning the size of the different terms in the potential, it is 
remarkable how, for small (big) values of the $D$ ($B$) parameter
it is the $F_{\varphi}$ term the one that dominates over $F_S$ in the
potential and therefore cancels the $3|W|^2$. Given that $K'<18$
that means that in this region of the parameter space (see the 
discussion after Eq.~(\ref{Kp18})) 
\be
\xi^2 = \frac{3}{4} \left( 1 - 
 \sqrt{1 - \frac{4}{3} \frac{\qq}{N_c}} \right)^{2} \;\;.
\label{xim}
\ee
As $B$ decreases (i.e. $D$ increases) the size of both $F$ terms becomes 
comparable and, eventually, the value of $\xi^2$ tends to a constant 
value given by $(q+\bar{q})/N_c$.
This can be understood from Eq.~(\ref{ours}) in the asymptotic regime of
``small'' $B$ (see Fig.~1). 
Defining  $\varepsilon \equiv 
e^{-B(\sqrt{{\rm Re} S_{min}}-\sqrt{S_0})}$ we have in this regime
\br
K_{np}|_{min} & \sim & \frac{D \varepsilon}{B \sqrt{{\rm Re} S_{min}}}  
\;\;, \nonumber \\
K'_{np}|_{min} & \sim & \frac{-D \varepsilon}{{\rm Re} S_{min}}  
\;\;, \label{kpmin} \\
K''_{np}|_{min} & \sim & \frac{B D \varepsilon}{({\rm Re} S_{min})^{3/2}}  
\;\;, \nonumber 
\er
where the subscript $min$ denotes values at the minimum.  
We can now evaluate  $f(\xi^2)$, which is given by
\be
f(\xi^2) = 1- \frac{2K'}{\dgs K''} \sim 1 + \frac{8 
\sqrt{{\rm Re} S_{min}}}{\dgs B}\ .
\ee
Imposing the limit of Eq.~(\ref{qbound}),
$f(\xi^2) < 3\N/[4(\qq)]$, we get a lower
bound on $B$, which is the one shown in Fig.~1 for each example,
\be\label{lowbbound}
B \geq  \frac{8 \sqrt{{\rm Re} S_{min}}}{\delta_{GS} 
\left( \frac{3 N_c}{4 (\qq)} -1 \right) }
\sim
\frac{96 \sqrt{{\rm Re} S_{min}}}{1- 12\dgs} \;\;.
\ee
It can also be seen in Fig.~1 that $B$ approaches asymptotically 
this bound, and
therefore $f(\xi^2)$ will tend to a constant value in this limit.
Using Eq.~(\ref{fsffi2}) we finally obtain that $\xi^2$ goes to an 
asymptotic value given by $(\qq)/\N$.
Using Eq.~(\ref{mDmF}) this means that the ratio $q_{\phi} m_F/m_D$ 
tends to $1$.
In fact, since $|K'|<18$ in these particular models, this same equation
tells us that $q_{\phi} m_F/m_D$ approaches $1$ from below.
In other words, we are in
a scenario where the scalar soft masses are dominated by the 
$D$-term (see the discussion at the end of section III).
All this is illustrated in 
Fig.~2 for the same three cases presented in Fig.~1. 
%%%%%%%%%%%%%%%%%%%%%%%%%%%figure%%%%%%%%%%%%%%%%%%%%%%%%%%%%%
\begin{figure}
\centerline{  
\includegraphics[height=9cm]{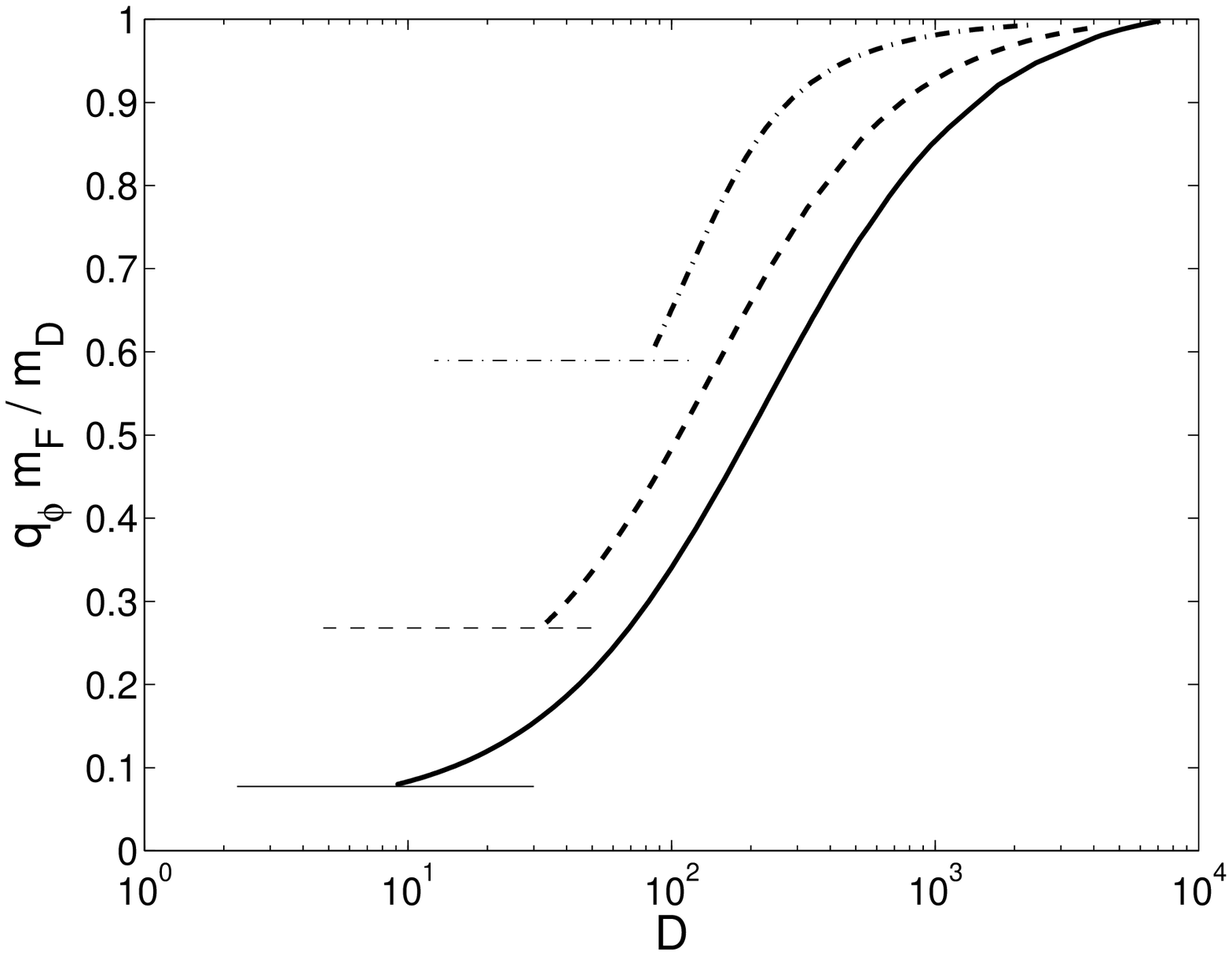}
}
\caption{}
{\footnotesize Plot of $q_{\phi} m_F/m_D$ vs $D$, with $B$ fixed to 
the values given by the previous figure. Same cases as before. The
thin lines show the corresponding lower bound on $q_{\phi} m_F/m_D$ 
derived from Eq.~(\ref{qbound}).}
\end{figure}
%%%%%%%%%%%%%%%%%%%%%%%%figure%%%%%%%%%%%%%%%%%%%%%%%%
The left hand limit of each curve is determined by the 
corresponding value of $\xi^2$ given by Eq.~(\ref{xim}), that is 
corresponds to $V_0=0$ driven by $F_\varphi$, whereas as we move in
the parameter space towards larger (smaller) values of $D$ ($B$)
the $F_S$ term contributes more to the cancellation of the 
cosmological constant until we reach the constant value defined by
$\xi^2=(\qq)/\N$. Moreover the ratio $q_{\phi} m_F/m_D$ tends
to its maximum value.
Therefore this is an ansatz for $K_{np}$ for which
we get the different possibilities for cancelling the cosmological 
constant, generically with the soft scalar masses dominated by their $D$ 
contribution.

\section{Conclusions}

Four-dimensional superstring constructions frequently present
an anomalous $U(1)_X$,
which induces a Fayet-Iliopoulos (F-I) D-term. On the other hand
SUSY breaking through gaugino condensation is a desirable feature
of string scenarios.  The interplay between these two facts is not
trivial and has important phenomenological consequences, especially
concerning the size and type of the soft breaking terms.
Previous analyses \cite{binet96,arkan98,dvali98},
based on a global SUSY picture,
led to contradictory results regarding the relative contribution
of the F and D terms to the soft terms.

In this paper we have examined this issue, generalizing the  
work of ref.~\cite{arkan98} to SUGRA, which is the
appropriate framework in which to deal with effective theories
coming from strings. We have extended the formulation of the F-I 
term made in \cite{arkan98} to the SUGRA case, as well
as considered the complete SUGRA potential in the analysis.
This allows to properly implement the cancellation of the cosmological
constant (something impossible in global SUSY), which is crucial
for a correct treatment of the soft breaking terms.
Moreover this condition yields powerful constraints on the K\"ahler 
potential, which leads to definite predictions
on the relative size of the contributions to the soft terms. 
In particular we obtain the following hierarchy of masses
\be
\label{hierarchy2}
m_{\lambda}^2 \ll m_F^2 < m_D^2 \;\;,
\ee
where $m_F^2$, $m_D^2$
are the contributions from the F and D terms, respectively, to the  
soft scalar masses, and $m_{\lambda}^2$ are the gaugino masses.
These results amend those obtained
in ref.~\cite{arkan98}. The last inequality could be reversed
if $K'$, i.e. the derivative of the K\"ahler potential
with respect to the dilaton at the minimum, is large ($\simgt 20$).
This is not surprising, since in this way
the F-I scale (which is proportional to $K'$) can be made
comparable to  $M_P$, so that the role of gravity as the
messenger of the (F-type) SUSY breaking is not overriden
by the ``gauge mediated'' (D-type) SUSY breaking associated 
to the F-I term. This situation cannot be excluded,
but in our opinion it is unlikely to happen. The first inequality in
(\ref{hierarchy2}) is rather worrying since it   
poses a problem of naturality.
Namely, given that gaugino masses must be compatible with their 
experimental
limits, the scalar masses must be much higher than 1 TeV, leading
to unnatural electroweak breaking. This conclusion seems inescapable
in this context.
   
Finally, we have illustrated all our results with explicit examples,
in which the dilaton is stabilized by a gaugino condensate and
non-perturbative corrections to the K\"ahler
potential, keeping a vanishing cosmological constant. There it is 
shown how the various contributions to the soft terms are in agreement
to the hierarchy expressed in Eq.~(\ref{hierarchy2}).

\section*{Acknowledgements}
JAC and JMM thank Michael Dine and  Steve Martin  for useful discussions
held at the University of Santa Cruz, and the NATO project CRG 971643 that
made it possible.
The work of TB was supported by JNICT (Portugal), BdC was supported 
by PPARC and JAC and JMM were supported by CICYT of Spain 
(contract AEN95-0195). 
Finally, the authors would like 
to thank the British Council/Acciones Integradas program for the 
financial support received through the grant HB1997-0073.

\end{document}